\documentclass[aps,prl,reprint]{revtex4-2}
\usepackage{amsmath,amsfonts,amssymb}
\usepackage{graphicx}
\usepackage{hyperref}
\usepackage{booktabs}
\usepackage{tikz}
\usetikzlibrary{quantikz}
\usepackage{braket}
\usepackage{multirow}

\begin{document}

\title{Experimental Demonstration of the PBR Test on a Superconducting Processor}

\author{Songqinghao Yang}
\affiliation{Cavendish Laboratory, Department of Physics, University of Cambridge, Cambridge CB3 0HE, UK}

\author{Haomu Yuan}
\affiliation{Cavendish Laboratory, Department of Physics, University of Cambridge, Cambridge CB3 0HE, UK}

\author{Crispin H. W. Barnes}
\affiliation{Cavendish Laboratory, Department of Physics, University of Cambridge, Cambridge CB3 0HE, UK}

\date{\today}

\begin{abstract}
We present an experimental implementation of the Pusey--Barrett--Rudolph (PBR) no-go theorem on IBM’s 156-qubit Heron2 Marrakesh superconducting quantum processor. By preparing qubits in a set of non-orthogonal states and evolving them under carefully compiled unitary circuits, we test whether one can interpret the hidden variable model for quantum states as merely epistemic---reflecting ignorance about some underlying physical reality. To account for realistic hardware imperfections, we derive noise-aware error tolerance based on decoherence models calibrated to the device’s performance. Our results show that a significant majority of adjacent qubit pairs and adjacent five-qubit configurations yield outcome statistics that violate the epistemic bound, thus ruling out the \textit{epistemic} interpretation of quantum mechanics. Furthermore, we observe a clear trend: the probability of passing the PBR test decreases as the spatial separation within the quantum processor between qubits increases, highlighting the sensitivity of this protocol to connectivity and coherence in Noisy Intermediate-Scale Quantum (NISQ) systems. These results demonstrate the PBR test as a promising device-level benchmark for quantumness in the presence of realistic noise.

\end{abstract}

\maketitle

\section{Introduction}

Quantum states serve as fundamental mathematical objects in quantum theory, yet their physical interpretation remains deeply contested. Two principal perspectives dominate this discussion: the $\psi$-ontic view maintains that a pure quantum state directly corresponds to physical reality. In contrast, the $\psi$-epistemic position treats the state as representing knowledge about some underlying physical state \cite{leifer2014quantum}. This distinction becomes precise within ontological models, where preparing a quantum state $\ket{\psi}$ generates an ontic state $\lambda$ sampled from a distribution $\mu_\psi(\lambda)$. A model is $\psi$-ontic when any two distinct quantum states $\ket{\psi_0}$ and $\ket{\psi_1}$ correspond to distributions $\mu_0(\lambda)$ and $\mu_1(\lambda)$ with disjoint support, meaning $\lambda$ uniquely determines $|\psi\rangle$. Conversely, a $\psi$-epistemic model permits some quantum states to share overlapping distributions, allowing a single $\lambda$ to correspond to multiple quantum states.

\begin{figure}
    \centering
    \includegraphics[width=1\linewidth]{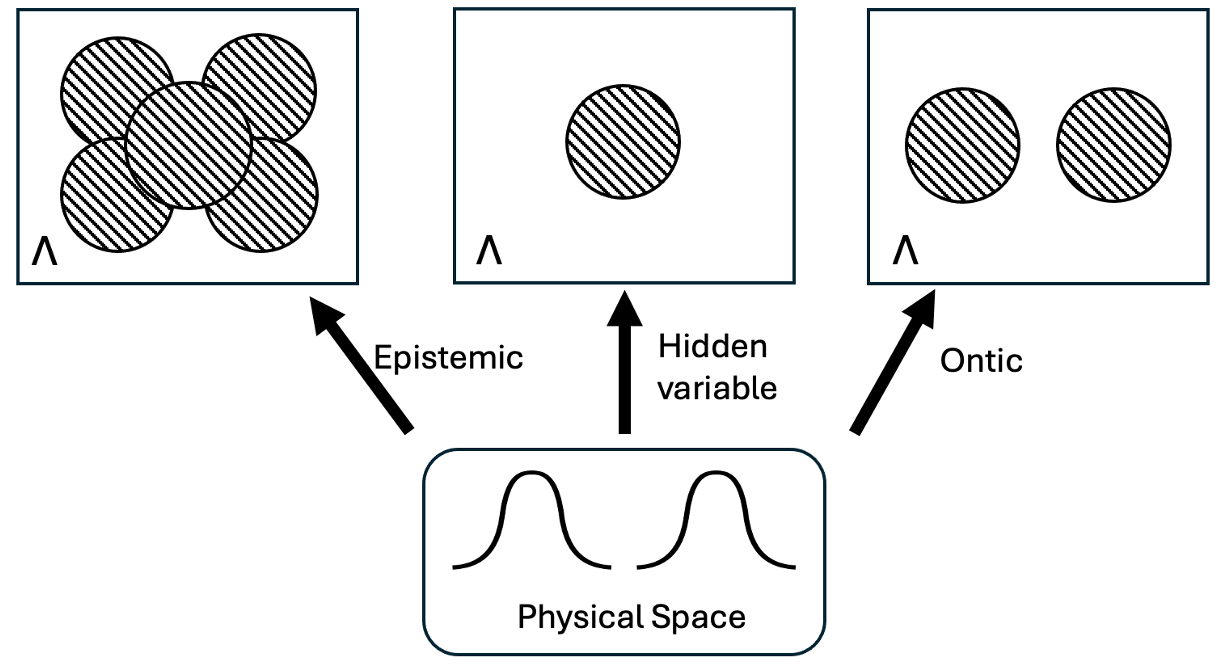}
    \caption{Illustration of three interpretations of the quantum states in the physical state (bottom) and the ontic space $\Lambda$: (Top left) \textit{Epistemic}—different quantum states correspond to overlapping distributions over $\Lambda$---implying $\psi$ reflects knowledge---ruled out by the PBR theorem. (Top middle) \textit{Hidden variable}—the same ontic state consisting of multiple physical states, suggesting incomplete knowledge of the quantum state, thus the need for a hidden variable model. (Top right) \textit{Ontic}—distinct quantum states map to disjoint state, implying representation of physical reality.
}
    \label{fig:intro}
\end{figure}

The Pusey-Barrett-Rudolph (PBR) theorem provides a crucial no-go result in this context \citep{pusey2012reality}. Under the assumption of preparation independence---independently prepared systems have product ontic states---the theorem establishes that any viable hidden-variable model must not be $\psi$-epistemic. Mathematically, preparation independence requires the joint ontic distribution to factorize as $\mu_{1,2}(\lambda_1,\lambda_2) = \mu_{1}(\lambda_1)\mu_{2}(\lambda_2)$. The PBR theorem demonstrated that any overlap between $\mu_0$ and $\mu_1$ would enable measurements to produce outcomes forbidden by quantum mechanics (Fig.\ref{fig:intro}). The epistemic hidden variable model is thus ruled out by the PBR test, which bears similarity to Bell’s test, which ruled out the local hidden variable \cite{bell1964einstein}.

Many research works have been conducted to improve both the experimental and theoretical frameworks for the PBR test \cite{miller2013alternative, patra2013no, liao2016experimental, aaronson2013psi, leifer2013maximally}. The experimental test of the PBR theorem involves preparing $n$ systems in one of two non-orthogonal states $|\psi_0\rangle$ or $|\psi_1\rangle$, followed by an entangling measurement designed such that each of its $2^n$ outcomes has zero probability for exactly one possible input configuration. In a $\psi$-epistemic model where the distributions $\mu_0$ and $\mu_1$ overlap with probability $\delta > 0$, there exists a finite probability $\delta^n > 0$ that all $n$ systems simultaneously occupy the same region. The overlap would make all measurement outcomes possible, directly contradicting quantum mechanical predictions where certain outcomes are strictly prohibited for specific inputs. The total variation distance $D(\mu_0,\mu_1) = \frac{1}{2}\int d\lambda|\mu_0(\lambda)-\mu_1(\lambda)|$ quantifies this distinction, with $D=1$ indicating disjoint distributions as required by $\psi$-ontic models \cite{patra2013experimental, barrett2014no, branciard2014psi}. 

While previous experimental tests of the PBR theorem have been successfully conducted using trapped ion systems~\cite{nigg2015can} as well as spin systems \cite{faroughi2021three}, our work represents the first experimental realization on a superconducting qubit architecture. We expanded the PBR protocol from the 2-qubit system size to a test on five qubits, using IBM’s 156-qubit Heron2 Marrakesh processor, significantly extending the scope of prior tests. We demonstrate that the protocol remains viable on a more scalable and commercially relevant quantum platform.

To support this scaling, we leverage the full spatial extent of the processor rather than limiting the experiment to locally connected qubits. By incorporating qubit pairs with varying physical distances, our implementation captures the effects of non-local connectivity and architectural inhomogeneity, providing a more realistic assessment of performance across the whole chip. Our approach also introduces a more detailed and comprehensive noise modelling framework. We combine a depolarizing error model—capturing gate-level imperfections—with a decoherence model based on calibrated $T_1$ and $T_2$ times. This dual-model strategy enables us to determine new error tolerance when the PBR test is expected to fail, revealing the tolerances required for successful protocol execution under realistic conditions. We further support the experimental results with analytical modelling that incorporates these noise sources to verify consistency between theoretical predictions and observed data. The robust formulation of the PBR theorem, which allows for minor deviations from ideal outcome probabilities, enables meaningful interpretation of experimental results even in the presence of hardware imperfections. 

Our results reveal tension with $\psi$-epistemic models adhering to preparation independence \cite{hall2011generalisations}, though the conclusive rejection of such models remains contingent on the noise characteristics of the system. The experimental outcomes for $n=2$ cases show strong contradiction with $\psi$-epistemic predictions within error margins. At the same time, the $n=5$ implementations highlight the sensitivity of the PBR test to decoherence and gate errors—with success rates decaying predictably with circuit depth. Similar results can be concluded from the statistics of the PBR test on distant qubit pairs, where the decoherence leads us to the epstemic interpretation of quantum mechanics as the circuit depth and width increases. This noise dependence suggests the PBR protocol could serve as a quantitative benchmark for quantum processors: the observed deviation from ideal predictions correlates directly with the device’s effective error rates. 

Our simulated experiments reveal a discrepancy, to some extent, between predicted and observed results as the system scales. While noise modelling based on calibrated $T_1$, $T_2$, and depolarizing errors provides good agreement in the small-$n$ regime, it becomes insufficient at larger scales, indicating an incompleteness in current noise models. This gap highlights the need for a more precise characterization of correlated and non-Markovian noise processes to accurately simulate and interpret complex quantum interference phenomena in larger systems.

\section{Theory and Noise Bounds}
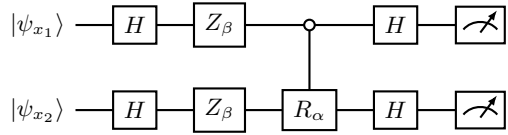
\begin{figure}[htb!]
\centering
\begin{quantikz}
\lstick{$|\psi_{x_1}\rangle$} & \gate{H} & \gate{Z_\beta} & \octrl{1} & \gate{H} & \meter{} \\
\lstick{$|\psi_{x_2}\rangle$} & \gate{H} & \gate{Z_\beta}          & \gate{R_\alpha}       & \gate{H} & \meter{}
\end{quantikz}
\caption{The general PBR test circuit for $n=2$. Input qubits are prepared in either $|\psi_0\rangle$ or $|\psi_1\rangle$. The entangling measurement includes $Z_\beta = R_z(\beta)$ and $CR_\alpha = CR_z(\alpha)$ rotations conditioned on the `0’ state and final Hadamards before measurement. Each measurement outcome $k$ is guaranteed (by the choice of $\alpha,\beta$) to have zero probability on one of the four input states. In the presence of noise, `forbidden’ outcomes occur with a small probability.}
\label{fig:pbr_circuit}
\end{figure}

\paragraph*{Ideal PBR measurement:}  We consider two pure single-qubit states $\ket{\psi_0}=R_Y(\theta)\ket{0},\ket{\psi_1}=R_Y(-\theta)\ket{0}$ with non-vanishing overlap. In the $n=2$ test, two qubits are prepared independently in  $\ket{\psi_{x_j}}$ with $\theta=\frac{\pi}{4}$, indexed by binary string $x_j \in \{0,1\}$, and $\ket{\psi_0}=\cos(\frac{\pi}{8})\ket{0} + \sin(\frac{\pi}{8})\ket{1}, \ket{\psi_1}=\cos(\frac{\pi}{8})\ket{0} - \sin(\frac{\pi}{8})\ket{1}$. We then apply an entangling circuit parameterized by angles $(\alpha,\beta)$---see Fig.~\ref{fig:pbr_circuit} for details. One can parametrize the PBR circuit with certain $(\alpha,\beta)$ so that the measurement has zero probability of obtaining outcomes $x_1x_2$, with the initial state prepared independently in the product state $\ket{\psi^{(0)}_0} \otimes \ket{\psi^{(1)}_1}$, where the superscript indicates the qubit index. For instance, outcome $00$ should be impossible for $\ket{\psi^{(0)}_0\psi^{(1)}_0}$, outcome $01$ impossible for $\ket{\psi^{(0)}_0\psi^{(1)}_1}$, etc.  Thus, in the ideal case, the observed frequency of each `forbidden’ outcome is zero if one prepares the corresponding input state.

In general, for a $n$-qubit system, $|\Psi(x_1\ldots x_n)\rangle=|\psi^{(1)}_{x_1}\rangle\otimes\cdots\otimes|\psi^{(n)}_{x_n}\rangle$, there are $2^n$ possible input strings $x_1\ldots x_n$. The extension of $n$-qubit PBR entangling circuit forbids the string $x_1\ldots x_n$ measured.

\paragraph*{Overlap and distance:} To quantify the noise tolerance in state preparation, we use the classical trace distance $D(\mu_0,\mu_1)$ between $\mu_0(\lambda),\mu_1(\lambda)$.  If the experiment were ideal, then observing zero appearances for the forbidden outcomes would imply $D(\mu_0,\mu_1)=1$---representing disjoint ontic supports. With finite noise, however, if the probability of each forbidden outcome is bounded by $\epsilon_{\rm tol}$, then 
\begin{equation}
    D(\mu_0,\mu_1)\;\ge\;1 - 2\epsilon_{\rm tol}^{\frac{1}{n}}.
    \label{eq:pbr_bound}
\end{equation}
Following~\cite{maroney2012statistical, nigg2015can}, we consider to rule out the maximally $\psi$-epistemic model in this work. Therefore, the classical trace distance is the natural choice of state preparation metric to characterize the probability distribution of the noisy PBR test circuit. The quantum trace distance for two pure states is:
\begin{equation}\label{eq:quantum_trace}
D_Q\bigl(\ket{\psi_0},\ket{\psi_1}\bigr)
:= \sqrt{1 - |\braket{\psi_0|\psi_1}|^2}.
\end{equation}
And it is not hard to derive the inequality relation
\[
D(\mu_0,\mu_1) \geq D_Q(\ket{\psi_0},\ket{\psi_1}),
\]
which means that for all pairs of states the classical trace distance suffices for the noise tolerance equality to hold. By demonstrating that the measured error probabilities fall strictly below the classical overlap bound, we expose a gap that purely epistemic explanations cannot bridge. We test the device by estimating the forbidden-outcome probabilities $p_k$ for each input. Passing the PBR test can be taken to mean that $p_k$ is low enough that the inferred $D(\mu_0,\mu_1)$ is close to 1. We will compare the observed $p_k$ to the bound
\begin{equation}
    \epsilon_{\rm tol} = \frac{(1-D\bigl(\mu_0,\mu_1\bigr))^n}{2^n},
    \label{bound}
\end{equation}
implied by Eq.\eqref{eq:pbr_bound} and Eq.\eqref{eq:quantum_trace} in the maximally $\psi$-epistemic model. The bound can be directly modelled with the overlap between the two non-orthogonal initial states that we choose. In the noisy setting, this changes to
\begin{equation}
    \epsilon^{\rm noise}_{\rm tol} = \frac{(1-D\bigl(\mu'_0,\mu'_1\bigr))^n}{2^n},
    \label{noise_bound}
\end{equation}
where $\mu'_0,\mu'_1$ are the ontic distribution for the independent states $\ket{\psi_0’}$ and $\ket{\psi_1’}$, prepared on superconducting quantum computer with noisy unitary operations.
\paragraph*{Depolarizing model:}  We model gate errors by independent depolarizing channels. Depolarising noise is a common abstraction in quantum error correction modelling, representing a uniform loss of coherence. For a single qubit state $\rho$, the depolarizing channel evolves the state into the linear combination of the maximally mixed state, $\frac{I}{2}$, with probability $p$ and the state itself with probability $1-p$.
Mathematically, the channel is given by
\begin{equation}
    \rho \mapsto (1 - p)\rho + \frac{p}{2}I.
\end{equation}
In our modelling, each single-qubit gate is followed by a depolarizing channel with probability $p_1$, and each two-qubit gate is followed by a two-qubit depolarizing channel with probability $p_2$, using values extracted from the processor’s daily calibration data. This simplification allows us to estimate the leakage of quantum information into noise analytically and bound the likelihood of nominally forbidden measurement outcomes.
For instance, in a circuit containing $G_1$ one-qubit gates and $G_2$ two-qubit gates, then the synthesize error of the depolarising noise is: 
\begin{equation}
\epsilon_{\rm dep} \approx1 - (1-p_1)^{G_1}(1-p_2)^{G_2}.
\label{eq:depol}
\end{equation}
Since the error tolerance is only dependent on the overlap between initial states, which is generated by a single $R_Y(\theta)$ rotation gate, with the ideal quantum trace distance being $D_Q = \sqrt{1 - |\braket{\psi_0|\psi_1}|^2}=\rm sin(\theta)$. After each rotation, we apply a depolarizing channel
\begin{align}
    \rho_{i={0,1}}^{\rm noisy} &= (1-\epsilon_{\rm dep})\rho_{i={0,1}}^{\rm ideal} + \epsilon_{\rm dep}\,\frac{I}{2},
\end{align}
where $\epsilon_{\rm dep} \approx 1 - (1-p_1)^{G_1=1}(1-p_2)^{G_2=0}.$
The overlap of the two noisy states is
\begin{align*}
D_Q(\ket{\psi_0’},\ket{\psi_1’})=& (1-\epsilon_{\rm dep})D_Q(\ket{\psi_0},\ket{\psi_1}) \\ 
 = &(1-\epsilon_{\rm dep})\rm sin(\theta).
\end{align*}
Similarly, the simulated state vector evolution of the multi-qubit PBR circuit can be estimated analytically using our noise model. 

\paragraph*{Thermodynamical model:}  We also simulate the noise by amplitude damping and dephasing channels based on time parameters $T_1, T_2$, respectively. 
Given the duration of implementing one qubit gate by $t_g$ in the quantum device, each qubit’s amplitude-damping probability per gate is $p_{\rm ad}=1-\exp(-t_g/T_1)$, and pure-dephasing $p_\phi=1-\exp(-t_g/T_2)$. 
The worst-case probability that any one of $n$ qubits decays in one gate is $1-(1-p_{\rm ad})^n\approx n\,p_{\rm ad}$ for small $p_{\rm ad}$.  Over the circuit, with multiple gates sequentially on each qubit, the cumulative error scales roughly as $n(1-e^{-N t_g/T_1})$, where $N$ is the total gate count. For simplicity, we approximate the dominant error contribution by 
\begin{equation}
\epsilon_{\rm dec}\;\approx\;n\bigl(p_{\rm ad}+p_\phi\bigr)\,,
\label{eq:decoh}
\end{equation}
taking a single effective gate duration per qubit. For example, with $T_1=192~\mu$s, $t_g\sim36$\,ns (single-qubit) and $\sim68$\,ns (two-qubit), we have $\bar{p}_{\rm ad}\approx2.8\times10^{-4}$ $\bar{p}_{\rm ad}\approx2.8\times10^{-4}$, and if $T_2=95~\mu$s, then $\bar{p}_\phi\approx5.6\times10^{-4}$, giving $\bar{p}_{\rm ad}+\bar{p}_\phi\approx8.4\times10^{-4}$, where $\bar{p}$ denotes the average noise over the single and two-qubit unitaries. For $n=5$ qubits this yields a deviation $\epsilon_{\rm dec}\approx5\times8.4\times10^{-4}\approx0.4\%$.  This analytical error bound accounts for both local decoherence and crosstalk during the gate sequence. Note that we neglect additional errors from idling (Identity gates). Within the thermodynamical model, we also consider the readout duration for decoherence in the thermodynamical model, which is typically in the order of a few $\mu s$.

\paragraph*{Readout error:} In the final stage of the simulation, we account for classical readout errors, which are described by the conditional probabilities \(P(A \!\mid\! B)\), where \(A \in \{0,1\}\) is the measured bit value and \(B \in \{0,1\}\) is the actual outcome of the quantum measurement. For a single qubit, these probabilities form the readout-error matrix
\begin{equation}
M = \begin{pmatrix}
P(0\!\mid\!0) & P(0\!\mid\!1)\\
P(1\!\mid\!0) & P(1\!\mid\!1)
\end{pmatrix}
=
\begin{pmatrix}
1 - p_{10} & p_{01}\\
p_{10}     & 1 - p_{01}
\end{pmatrix},
\end{equation}
which maps the ideal measurement probabilities to the experimentally observed outcome distribution and the off-diagonal elements capture the bit-flip error rates: \(p_{10} = P(1 \!\mid\! 0)\) represents the error probability of flipping the measure of \(\ket{0}\) to `1' and \(p_{01} = P(0 \!\mid\! 1)\) is the probability of flipping the measure of \(\ket{1}\) to `0'.

Thus, implementing these noise models gives an approximation of the  error bound [Eq.~\ref{noise_bound}] for the probability of the forbidden outcomes, i.e. Eq.~\ref{eq:depol} and Eq.~\ref{eq:decoh}. In the next section, we will compare the actual measurements to these theoretical $\epsilon^{\rm noise}_{\rm tol}$ values.  

\section{Experimental Implementation}
We used IBM’s Heron2 Marrakesh 156-qubit processor. The native gates include \texttt{CZ}, \texttt{SX}, \texttt{X}, \texttt{RZ}, \texttt{RX}, \texttt{ID}, and \texttt{RZZ}. Circuit compilation used Qiskit’s transpiler targeting nearest-neighbour connectivity. Calibrated gate error rates, readout errors, and $T_1$, $T_2$ values were obtained on the day of execution. Each $n=2$ and $n=5$ circuit was executed with  100,000 shots to ensure statistical convergence of output frequencies.

\noindent\textbf{Choice of preparation angle $\theta$:} In the original PBR construction \cite{pusey2012reality}, the number of subsystems $n$ must be large enough so that the pairwise overlap between the quantum states, quantified by $|\braket{\psi_0|\psi_1}|^2 = \cos^2\theta$, satisfies the threshold condition:
\begin{equation}
\frac{\pi}{2} \geq \theta \geq 2\arctan\left(2^{1/n} - 1\right), \label{eq:pbr_theta}
\end{equation}
which ensures that each outcome in the joint measurement has strictly zero probability under at least one of the \(2^n\) preparation scenarios. The circuit parameter $\alpha$, $\beta$ are accordingly chosen by solving the equation
\begin{equation}
e^{i\alpha} + \left(1+e^{i\beta}\tan\frac\theta2\right)^n - 1 = 0.
\end{equation}
For the $n=2$ case, the minimum valid value satisfying Eq.~\ref{eq:pbr_theta} is $\theta = \pi/4$. We adopt this smallest valid angle to simplify the circuit structure—since these \(\theta\) values yield shorter circuit depth, which is more robust to decoherence and gate noise. Specifically, we prepare $\ket{\psi_0} = \cos(\frac{\pi}{8})\ket{0} + \sin(\frac{\pi}{8})\ket{1}$ and $\ket{\psi_1} = \cos(\frac{\pi}{8})\ket{0} - \sin(\frac{\pi}{8})\ket{1}$, and select $\alpha=\pi$, $\beta=0$.

To further explore the robustness and generality of the test, we also perform additional experimental runs with enlarged angles, $1.2\times$, and the minimal value (i.e., $\theta=0.3\pi$). These tests allow us to validate the predictions of the PBR bound beyond the minimal circuit depth and to assess the sensitivity of forbidden-outcome statistics to increased state distinguishability. For larger \(\theta\), preparation becomes more distinguishable, but circuit depth increases—compounding the effect of gate errors and decoherence. As expected, we observe increased success probabilities for forbidden outcomes with larger \(\theta\), accompanied by slightly higher overall noise sensitivity in $n=5$ cases.

\begin{table}[htb!]
\centering
\caption{Average error tolerance thresholds for the PBR test under two noise models.}
\begin{tabular}{ccc}
\toprule
\textbf{Qubit number} & \textbf{Noise Model} & \textbf{Tolerance} \\
\midrule
n=2 
  & Thermodynamical & $2.72\pm 0.36\%$ \\
  & Depolarizing    & $2.14\pm 0.10\%$ \\
\midrule
n=5
  & Thermodynamical & $0.63\pm 0.41\%$ \\
  & Depolarizing    & $0.55\pm 0.09\%$ \\
\bottomrule
\end{tabular}
\label{tab:tolerance_summary}
\end{table}

\begin{figure}
    \centering
    \includegraphics[width=0.9\linewidth]{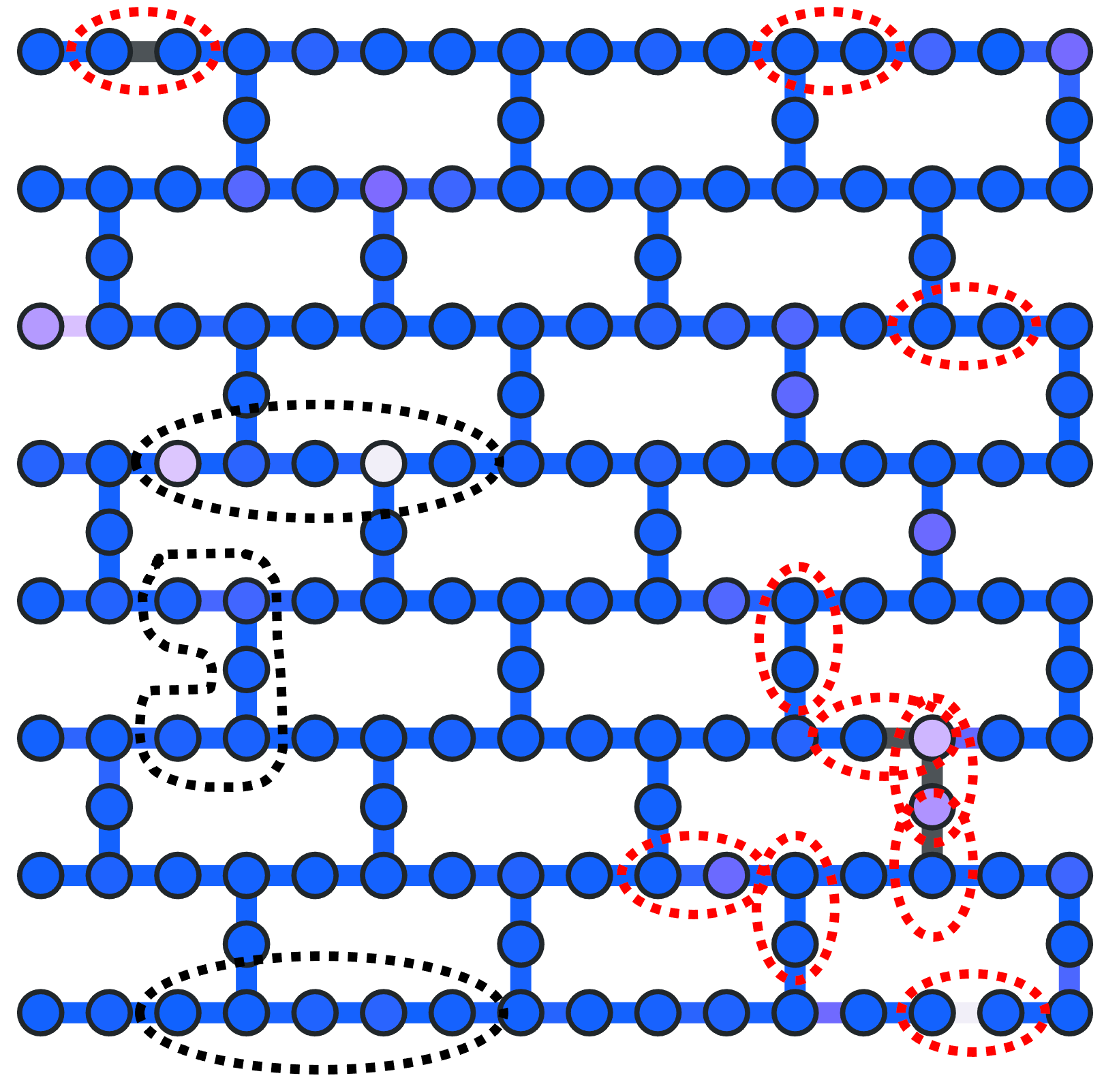}
    \caption{The layout of qubits on the Marrakesh processor. Each edge represents a physical two-qubit connection. Red circles indicate the adjacent two-qubit pairs test, while the black circles indicate those for an adjacent five-qubit test. The spatial distribution reflects how connectivity and localized noise affect the test outcome. Note that not all qubit pairs are shown in the figure for a clear visualization.}
    \label{fig:qubit_pair}
\end{figure}

\noindent\textbf{Adjacent qubit pairs ($n=2$):} We selected 20 distinct qubit pairs that are directly connected via tunable couplers (see the red circle in Fig.~\ref{fig:qubit_pair}), thus there will be no crosstalk between other qubits when implementing the PBR test. These included the 10 pairs with the highest fidelity \texttt{CZ} gates and the 10 with the lowest fidelity. Among these 20 experiments ($20 \times 2^2$ (variantion of initial state)=80 different circuits in total), approximately $97.5\%$ exhibited a violation of the $\psi$-epistemic bound---that is, the observed forbidden-outcome probabilities were low enough to satisfy the PBR criterion.

For instance, Qubit 1 ($T_1 = 173~\mu$s, $T_2 = 172~\mu$s and single unitary error around $2.1\times 10^{-4}$) paired with Qubit 2 ($T_1 = 239~\mu$s, $T_2 = 276~\mu$s and single unitary error around $2.8\times 10^{-4}$) have a two-qubit \texttt{CZ} gate error around $2.4 \times 10^{-3}$. Using these parameters in our noise model predicted forbidden-outcome probabilities on the order of $\sim 10^{-2}-10^{-3}$, and the observed forbidden-outcome rates $0.36_{-0.14}^{+0.14}\%$. The simulated error tolerance are summarized in Table.\ref{tab:tolerance_summary}.

However, as shown in the $\theta = 1.2\theta_0$ column of Table~\ref{table:pbr-summary}, the error rates for nearly all $n=2$ qubit pairs exceed the tolerance thresholds. This degradation is primarily due to the increased circuit complexity associated with random $\theta$.\footnote{Random here means that the value of $\theta$ could not be transpiled to a simple gate operation. For example, for a controlled rotation $\frac{\pi}{2}$ gates, it is equivalent to a $CZ$ gate. But for a random value, the controlled rotation gates have to be transpiled with multiple native gates.} Specifically, the PBR circuit for \(\theta > \theta_0\) involves additional arbitrary-angle single-qubit rotations and multi-controlled rotational gates, which are more susceptible to gate infidelity and transpilation overhead. In contrast, when using the minimal angle \(\theta = \theta_0 = \pi/4\), the circuit compiles more efficiently with fewer high-angle gates and exhibits greater resilience to noise. 

\noindent\textbf{Adjacent Five-qubit groups ($n=5$):} We selected 5 configurations of five qubits each to test the $n=5$ PBR circuit (see the black circle in Fig.~\ref{fig:qubit_pair}). These groups were chosen to explore different topologies: some formed a compact “corner” of the lattice, others a linear “chain,” some a high-coherence row, and one was a deliberately worse-case random set. The successful groups tended to have higher average $T_1, T_2$ and fewer two-qubit gates, as expected. Similar to the result for adjacent two-qubit pairs, the additional rotation gates for the $\theta=1.2\theta_0$ setup failed the test due to noise.

\begin{figure}
    \centering
    \includegraphics[width=0.9\linewidth]{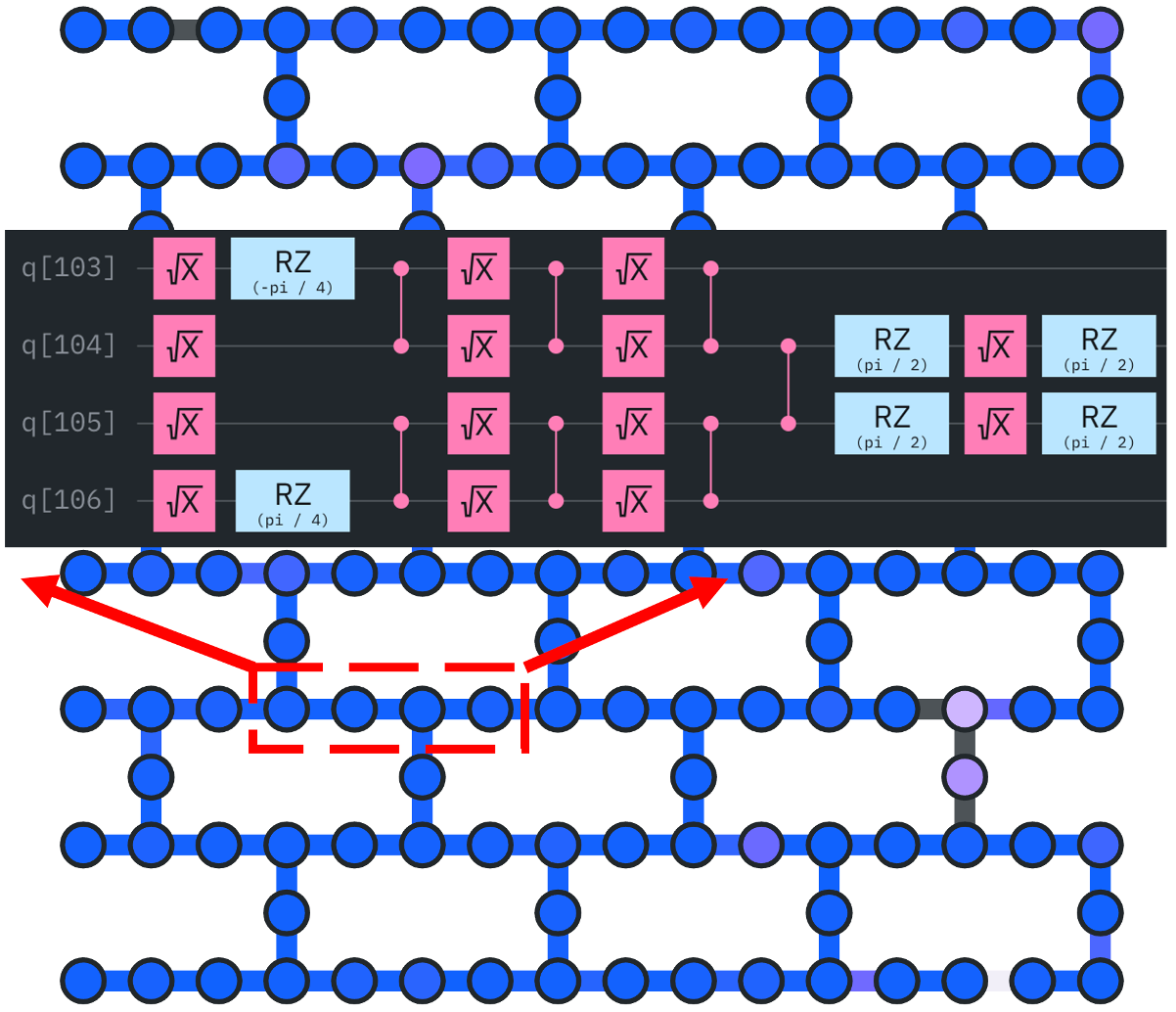}
    \caption{An example of a distance-2 two-qubit PBR test is shown here. The background illustrates the qubit layout of the Marrakesh architecture, with the four relevant qubits highlighted: [103, 104, 105, 106]. The inset displays the circuit automatically generated by Qiskit’s transpiler for this specific configuration.}
    \label{fig:transpile}
\end{figure}

\noindent\textbf{Distant qubit pairs ($n=2$):} To investigate the impact of qubit connectivity on the PBR test, we systematically increased the physical separation between qubit pairs—from nearest neighbors (with one intervening qubit) up to pairs separated by eight intermediate qubits. For instance, the transpiled quantum circuit for a distance-2 experiment is shown in Fig.~\ref{fig:transpile}. Although additional \texttt{SWAP} operations increased the circuit depth, non-adjacent pairs continued to pass the PBR test. However, the passing probability declined steadily---from 100\% at minimal separation to around 25\% at the maximum tested distance of seven intervening qubits (total qubit span $d=5$).

We observed a consistent trend: as the number of intervening qubits increased from 2 to 8, both the average error rate and worst-case performance worsened. This degradation is primarily due to the insertion of multiple \texttt{SWAP} gates, which increase circuit depth and expose the system to greater cumulative noise. As shown in Table~\ref{table:pbr-summary}, error rates rise from approximately $0.14\%$ at short distances to over $6\%$ at longer ones. Additionally, experiments with $\theta = 1.2\theta_0$ consistently performed worse, reflecting the sensitivity of the test to circuit complexity. These results underscore the combined influence of circuit structure and hardware topology on the reliability of foundational quantum protocols such as the PBR test.

Finally, we tested the most extreme case on the 156-qubit \texttt{Heron2} architecture: qubit-0 paired with qubit-155, spanning the entire device. This configuration failed the PBR test, yielding outcome statistics far outside the analytical tolerance bounds, with a standard deviation significantly beyond the acceptable limit. This result highlights the severe compounding effects of reduced connectivity and accumulated noise in large-scale quantum processors.
Since only one such full-span pair exists, best/worst statistics are not applicable (N/A).

\begin{table*}[htb!]
\centering
\caption{Summary of PBR test error rates on IBM Heron2 Marrakesh for different qubit configurations. Best and worst indicate the average operational (CZ gate) quality extremes.}
\begin{tabular}{lcccc}
\toprule
\textbf{Configuration} 
& \multicolumn{2}{c}{\(\theta = \theta_0\)} 
& \multicolumn{2}{c}{\(\theta = 1.2\theta_0\)} \\
\cmidrule(lr){2-3} \cmidrule(lr){4-5}
 & \textbf{Best (\%)} & \textbf{Worst (\%)} & \textbf{Best (\%)} & \textbf{Worst (\%)} \\
\midrule
$n=2$, adjacent pairs & $0.17 \pm 0.02$ & $0.56\pm 0.03$ & $2.19 \pm 0.16$ & $4.62\pm 1.15$ \\
$n=5$, adjacent chains & $0.32 \pm 0.10$ & $0.48\pm 0.18$ & $0.09 \pm 0.02$ & $0.10 \pm 0.02$  \\
\midrule
$n=2$, distant pair ($d=2$) & $0.14 \pm 0.04$ & $0.44\pm 0.04$ & $2.53 \pm 0.17$ & $5.39\pm 1.41$ \\
$n=2$, distant pair ($d=3$) & $0.44\pm 0.04$ & $0.45\pm 0.03$ & $2.62\pm 0.21$ & $5.39\pm 2.13$ \\
$n=2$, distant pair ($d=4$) & $0.67\pm 0.04$ & $1.49\pm 0.85$ & $2.63\pm 0.22$ & $5.64\pm 2.21$ \\
$n=2$, distant pair ($d=5$) & $0.69\pm 0.05$ & $5.09\pm 1.90$ & $2.64\pm 0.28$ & $5.71\pm 2.68$\\
$n=2$, distant pair ($d=6$) & $0.71\pm 0.05$ & $5.35\pm 2.45$ & $2.66 \pm 0.30$ & $6.08 \pm 3.19$ \\
$n=2$, distant pair ($d=7$) & $0.83\pm 0.09$ & $6.55\pm 2.59$ & $2.75 \pm 0.37$ & $7.16 \pm 3.43$ \\
$n=2$, distant pair ($d=8$) & $1.07\pm 0.24$ & $6.91\pm 3.08$ & $2.93 \pm 0.51$ & $7.69 \pm 3.60$ \\
$n=2$, full-span pair ($d=154$) & $6.30 \pm 0.65$ & N/A & $6.23 \pm 1.58$ & N/A \\
\bottomrule
\end{tabular}
\label{table:pbr-summary}
\end{table*}

All circuits were executed on real quantum hardware. Consequently, our noise modelling is based on the actual circuit elements after optimized transpilation to the chip’s native gate set \cite{zulehner2018efficient, murali2019noise} . For each input preparation, we recorded the frequency \( f_k \) of every measurement outcome \( k \). In particular, we focused on extracting the frequencies \( f_k \) corresponding to theoretically forbidden outcomes, which should ideally be zero in the absence of noise. A summary of the experimental results is provided in Table~\ref{table:pbr-summary}.

\noindent\textbf{Simulated Experiments:} We simulated the PBR circuits for two qubits with separations ranging from adjacent pair ($d=0$) to eight intervening qubits ($d=8$) using Marrakesh‐calibrated noise parameters. Under the thermodynamical noise model, the average PBR error increases from $6.25\%\pm1.08\%$ at $d=0$ to fluctuations of up to $6.80\%$ at larger separations, while under the depolarizing model the error spans from $6.23\%$ to $14.01\%$ across the same range. By comparison, the analytical tolerance thresholds for $n=2$ qubits are $2.72\%\pm0.36\%$ (thermodynamical) and $2.14\%\pm0.10\%$ (depolarizing) (Table~\ref{tab:tolerance_summary}), so although simulated errors agree within roughly $1\%$ for error tolerance calculation (see the example in \textit{Thermodynamical model} section), they diverge markedly as circuit depth grows—especially after transpilation inserts multiple SWAP gates—indicating that these simple noise models under‐estimate cumulative errors in deeper circuits. These baseline simulations underscore the need for more detailed, device‐informed noise modeling to accurately predict performance in large‐depth regimes.

\section{Discussion}
This work presents a noise‐aware implementation of the PBR no‐go test on IBM’s 156-qubit Heron2 superconducting processor, using two complementary error models calibrated to the device. Depolarizing noise is applied after each gate to capture control‐pulse imperfections, while amplitude‐ and phase‐damping channels during idling and readout model thermodynamic relaxation and dephasing. A large majority of both adjacent two- and five-qubit configurations exhibit forbidden-outcome frequencies that are below the limit restricted by purely epistemic (ignorance‐based) models, demonstrating that the observed error rates cannot be explained solely by ontological overlaps of underlying distributions. By comparing simulated outcome probabilities---including these noise operations and the readout matrix---to the measured data, we show that maintaining interference contrasts at or below our analytically derived tolerance is essential for passing the PBR test. The dependence of the passing probability on circuit depth, gate count, and qubit separation, therefore, provides a stringent, hardware-level benchmark of coherence and gate fidelity in NISQ devices.  

However, positioning the PBR protocol as complementary to the state-of-art metrics, such as Quantum Volume~\cite{cross2019validating} and cross‐entropy benchmarking~\cite{arute2019quantum}, would require further investigation. For example, the PBR test directly measures interference-based state indistinguishability under noise, whereas Quantum Volume captures holistic circuit performance, and cross-entropy quantifies sampling fidelity. By comparing our PBR-derived error tolerance with published Quantum Volume scores for similar hardware, readers can better appreciate the unique sensitivity of PBR to specific noise channels and its role in a diversified benchmarking suite.

Experimental PBR tests could help us understand the interpretation of different quantum mechanical models. From an epistemic perspective, decoherence is an informational process. Although the joint system--environment state $\ket{\Psi_{SE}}$ evolves unitarily, tracing out the environment yields a reduced density matrix
\begin{equation}
\rho_S = \mathrm{Tr}_E\left[ \ket{\Psi_{SE}}\bra{\Psi_{SE}} \right] \approx \sum_i p_i \ket{s_i}\bra{s_i},
\end{equation}
In which off-diagonal terms decay, and the hidden-variable distributions $\mu_0(\lambda)$ and $\mu_1(\lambda)$ increasingly overlap, driving the system toward classical probabilistic behaviour.

In contrast, the ontic interpretation treats the wavefunction as a real physical entity: each pure state $\ket{\psi}$ corresponds to a distribution $\mu(\lambda)$ with disjoint support. Unitary evolution, such as $\ket{\psi}_S \otimes \ket{E_0}_E \to \sum_i c_i \ket{s_i}_S \otimes \ket{E_i}_E$, redistributes coherence into orthogonal environmental subspaces without inducing overlap between previously disjoint ontic distributions. Nevertheless, as entanglement with the environment increases, even an ontic quantum processor gradually loses its advantage in interference. At this point, it begins to behave more like a classical stochastic system, marking the practical boundary between quantum and classical performance in NISQ devices. 

It is worth noting recent theoretical critique \cite{unnikrishnan2025information} of the PBR framework, arguing that the core PBR assumption---namely, that each quantum state corresponds to a fixed distribution of underlying ontic variables with strictly non‐overlapping support for orthogonal states---inevitably conflicts with the linear superposition principle and Born’s rule when one constructs superposed states. In particular, orthogonal superpositions forced to draw only from the original disjoint supports lead to logical contradictions, showing that PBR’s no‐go theorem rules out only this specific ontological model, not all $\psi$-epistemic interpretations. This perspective highlights that, even as our noise-aware implementation confirms the operational utility of the PBR test on NISQ hardware, alternative epistemic models may yet offer a consistent foundation for the wavefunction.

\section{Conclusion}
We have performed the first superconducting-qubit realization of the PBR theorem, applying it to 2- and 5-qubit circuits on IBM’s 156-qubit Heron2 hardware. We derive and verify noise-dependent bounds that quantify when an experimental violation excludes $\psi$-epistemic models. The results confirm quantum theory’s predictions under practical noise, supporting the ontic view of the wave function. Looking forward, extending PBR tests to larger $n$ or different state angles and using them to benchmark new devices could help understand noise impact to devices' `quantumness' and further define the practical quantum advantage.

\bibliographystyle{unsrt}
\bibliography{refs}

\end{document}